\input{aipcheck}

\documentclass[
  ,final            
  ]
  {aipproc}

\layoutstyle{6x9}

\begin{document}

\title{X-ray States of Black Hole Binaries in Outburst}


\author{R. A. Remillard}{
  address={MIT Kavli Institute for Astrophysics and Space Research}
}

\begin{abstract}
We continue to probe the properties of stellar-size black holes and
the physics of black-hole accretion using bright X-ray transients.
Progress has been made in the recognizing that the three states of
active accretion are related to different physical elements that may
contribute radiation: the accretion disk, a jet, and a compact corona.
Each of these states offers potential applications for investigation
via general relativity in the regime of strong gravity.
High-frequency QPOs are especially interesting in this regard, as the
evidence mounts for their interpretation as stationary 'voice-prints'
that may constrain black-hole mass and spin.
\end{abstract}

\maketitle


\section{Re-defining X-ray States}

It has been proposed recently that X-ray states of black hole binary
systems should be redefined in terms of quantitative criteria that
utilize both X-ray energy spectra and power density spectra (PDS)
\cite{mcc05}.  The goals of this effort are to capture the
essential elements of historical state descriptions, to incorporate
the complexity of black hole behavior revealed in extensive monitoring
campaigns with the {\it Rossi} X-ray Timing Explorer ({\it RXTE}), and to
correct misleading terminology.  

The redefinition of X-ray states utilizes four criteria: $f_{disk}$,
the ratio of the disk flux to the total flux (both unabsorbed) at 2-20
keV; the power-law photon index ($\Gamma$) at energies below any break
or cutoff; the integrated rms power in the PDS at 0.1--10 Hz ($r$;
expressed as a fraction of the average source count rate); and the
integrated rms amplitude ($a$) of a QPO detected in the range of
0.1--30 Hz.

It had been known for decades that the energy spectra of outbursting
black holes often exhibit composite spectra consisting of two
broadband components \cite{tan95}.  There is a multi-temperature
accretion disk \cite{sha73,mak86,gie04} with a characteristic temperature
near 1 keV.  This modified form of the black-body function is easy to
recognize.  Usually, it is seen in combination with a quasi-power-law
component, which may be modified by disk reflection \cite{don01,rev01}
and/or a cutoff energy (e.g. 100 keV) or a spectral break to a steeper
power law. Many black hole systems further exhibit a broad Fe emission
line (e.g. \cite{rey03,mil02}) and atomic absorption edges that are
primarily due to the cold ISM.  In some cases there are also
absorption features due to hot gas that is local to the binary system
(e.g. \cite{lee02}).

\begin{figure}
  \includegraphics[height=.65\textheight]{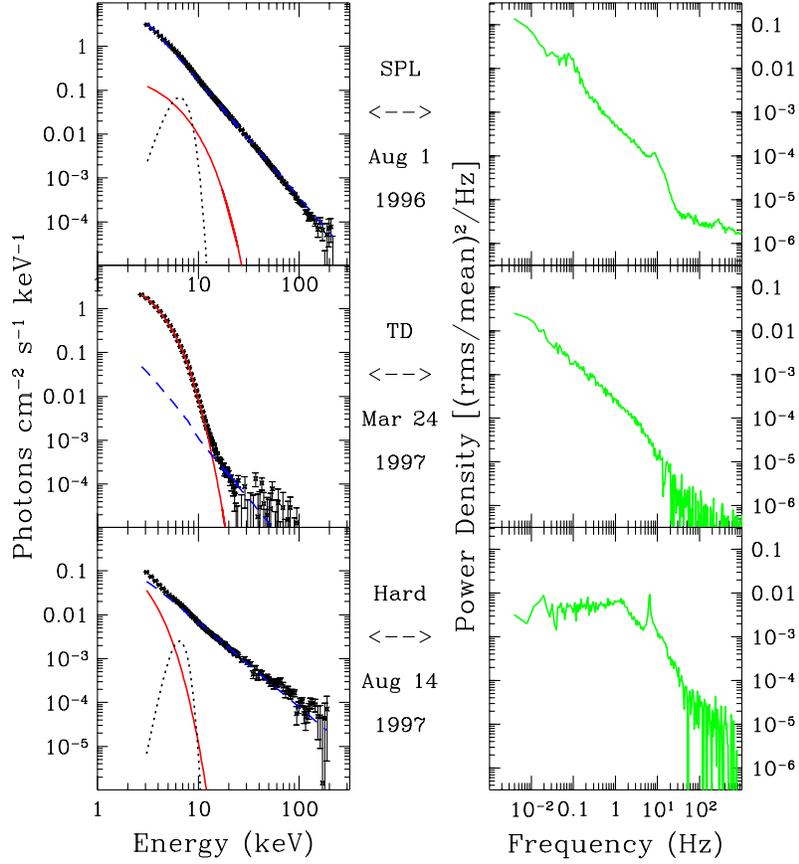}
  \caption{Examples of the 3 states of active accretion for the black
  hole binary GRO~J1655-40. Left panels show the energy spectra, with
  model components attributed to thermal-disk emission (red solid
  line), a power-law continuum (blue dashes) and a relativistically
  broadened Fe K--$alpha$ line (black dotted).  Power-law components
  for the SPL and hard states are distinguished by different values of
  the photon index (i.e. slope). The PDS (green solid lines) are shown
  in the right panels. A strong, band-limited continuum characterizes
  the hard state, while QPOs and the absence of the intense, broad
  continuum are usually seen in the SPL state.}
\end{figure}

There is a need to draw attention to those times when black hole
radiation is dominated by the heat from the inner accretion disk,
while there are no obvious temporal features that signify unexpected
oscillations or instabilities that complicate the picture.  Toward
this end, the themal state (formerly the ``high/soft'' state) is
defined by the following three conditions: (1) the disk contributes
more than 75\% of the total unabsorbed flux at 2--20 keV, i.e. $f >
0.75$, (2) there are no QPOs present with integrated amplitude above
0.5\% of the mean count rate, i.e. $a_{max} < 0.005$, and (3) the
integrated power continuum is low, with $r < 0.06$.

In principle, the normalization constant for the thermal component may
allow numerical estimates of the radius of the inner accretion disk,
if the source distance and disk inclination are accurately known
\cite{zha97a}. However, such estimates depend on disk models computed
under general relativity (GR), with careful attention to the inner
disk boundary condition and to effects of radiative transfer
\cite{mer00}.  Improved disk models may one day lead to trustworthy
results, and this could lead to measures of the dimensionless spin
parameter when the mass is well constrained via dynamical measures of
stellar motion in the binary system.

There are other occasions when the spectrum from an outbursting
black-hole system shows a much greater contribution from an X-ray
power-law component. Observations with {\it CGRO}--Ossie were
particularly valuable in showing that there were two forms of these
non-thermal spectra \cite{gro98}.  Spectral fits for the power-law
component yield clusters of $\Gamma$ values: one near 1.7 (hard state,
with an exponential decrease beyond $\sim 100$ keV) and the other near
2.5 (steep power law, with no apparent cutoff).  In each case, the
corresponding PDS also shows a distinct departure from the thermal
state. An illustration of the thermal state and the two non-thermal
states is given in Fig. 1.

In the hard X-ray state, the accretion-disk component is either
absent or it is modified in the sense of appearing comparatively 
cool and large.  The hard state has been clearly
associated with the presence of a steady type of radio jet
\cite{fen05,dha00}. Transitions to either the thermal state (e.g. GX339-4; 
\cite{fen99,cor00}) or an SPL state (e.g. Cyg X-1; \cite{zha97b}) 
effectively quenches the radio emission. This crucial, multi-frequency
advancement helps to demonstrate that the three X-ray states of black
hole binaries in outburst represent accretion systems that are very
different in terms of physical elements, geometry, and
energy-transport mechanisms. Hindsight analysis of X-ray data
correlated with radio signatures of the steady jet allows a definition
of the hard state, again based on three X-ray conditions: (1) $f <
0.2$, i.e. the power-law contributes at least 80\% of the unabsorbed
2--20 keV flux, (2) $1.5 < \Gamma < 2.1$ (for power-law, cutoff power
law, or broken power-law (using $\Gamma_1$), as appropriate), and (3)
the PDS yields $r > 0.1$.  

Current research on the hard state now addresses more detailed and
physical questions such as the jet energetics and ejection velocities
(\cite{fen05}; see also Gallo et~al., these proceedings), and the
whether the X-ray spectrum represents synchrotron radiation, inverse
Compton emission at the base of the jet, or a collimated outflow
related to an advection-dominated accretion flow (ADAF)
\cite{esi01,mar01,fro03,yua05}.

The steep power law (SPL) component was first linked to the power-law
``tail'' found in the thermal state, and it was widely interpreted as
inverse Compton radiation from a hot corona somehow coupled to the
accretion disk.  The picture became more complicated when X-ray QPOs
were first detected with Ginga for two sources: GX339-4 (6 Hz) and
X-ray Nova Muscae 1991 (3--8 Hz) \cite{miy91,miy93}. The QPOs, the
high luminosity, and the strength of the power-law component prompted
the interpretation that the QPOs signified a new black hole state,
labeled as the ``very high'' state. {\it RXTE} observations later 
showed that X-ray QPOs from black-hole binaries are much more common 
than had been realized \cite{vdk05}. 

As noted above, {\it CGRO} observations have shown that the SPL may
extend to photon energies as high as 800 keV \cite{gro98,tom99}). This
forces consideration of non-thermal Comptonization models
\cite{gie03,zdz04}. The QPOs impose additional requirements for an
oscillation mechanism that must be intimately tied to the electron
acceleration mechanism (in the inverse Compton scenario), since the
QPOs are fairly coherent ($\nu / \Delta \nu \sim 12$ ; \cite{rem02a})
and are strongest above 6 keV. Despite a wide range in SPL
luminosities, the SPL tend to dominate as the luminosity approaches
the Eddington limit (e.g. Fig. 1). Furthermore, the occasions of
high-frequency QPOs at 100-450 Hz in 7 black hole binaries almost
always coincide with a strong SPL spectrum \cite{mcc05}. Overall, the
many fundamental differences between the thermal and SPL properties
(see Fig. 1) force us to reject alternative state descriptions
that unify thermal and SPL observations under a single ``soft'' state.

The results from intense monitoring campaigns for black-hole
transients with {\it RXTE} motivated the redefinition of the very high
state as the ``SPL state'' \cite{mcc05}, with conditions: (1) $\Gamma
> 2.4$, (2) $r < 0.15$, and (3) either $f < 0.8$ while a QPO (0.1 to
30 Hz) is present in the PDS (with $a > 0.01$), \textbf{or} $f < 0.5$
with no QPOs (i.e. the disk contributes less than half of the flux).
Generally, the accretion disk remains visible in the X-ray spectrum in
the SPL state.  There may be modifications to the thermal parameters
($T_{in}, R_{in}$) during SPL episodes at high luminosity.  In such
cases, the disk appears unusually small and hot, and this is a likely
artifact of radiative transfer effects that occur when the disk is
viewed through a compact corona with moderate optical depth \cite{kub04}.

\begin{figure}
  \includegraphics[height=.75\textheight]{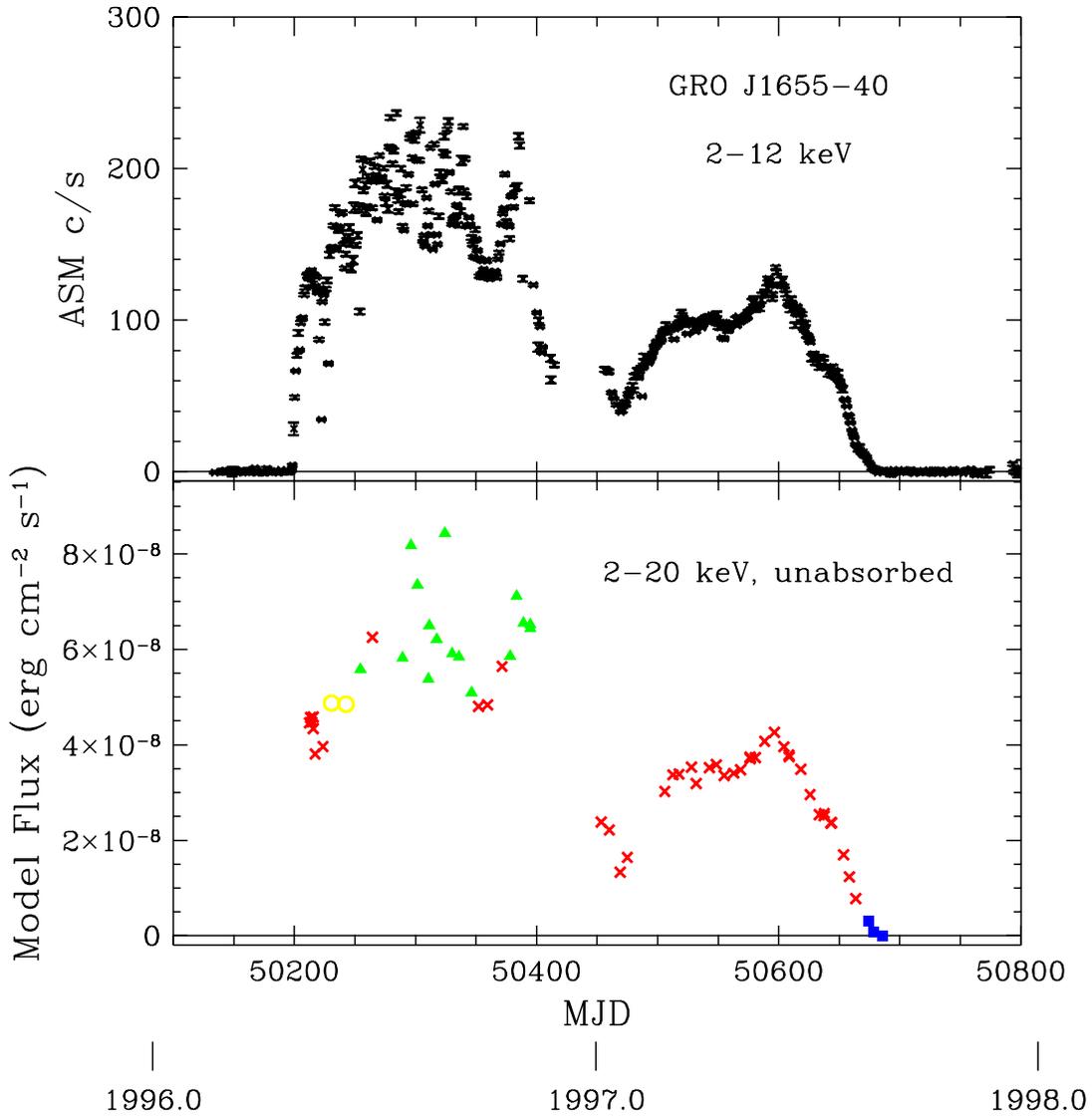}
  \caption{X-ray state evolution during the 1996-1997 outburst of
  GRO~J1655-40. The top panel shows the ASM light curve.  The bottom
  panel shows the model flux (2--20 keV, unabsorbed) from PCA pointed
  observations.  Here the symbol type denoted the X-ray state: thermal
  (red x), hard (blue square), steep power-law (green triangle), and
  any type of intermediate state (yellow circle). }
\end{figure}

This quantitative framework for the three states of active accretion
in black-hole binary systems is intended to tie X-ray states to each
of the three broad-band spectral components that have a distinct
physical origin and/or radiation mechanism. These states appear to
demonstrate a capacity for quasi-stability, but observations exhibit
their inherently transient nature.  Clearly, the parameter ranges
chosen for the three X-ray states leave substantial room for
intermediate conditions. Conversely, we abandon the effort to
pigeon-hole every observation into a well-defined state ; this
perspective merely honors the complexity of black hole outbursts as
seen in the {\it RXTE} era.

\section{Temporal Evolution of X-ray States}

The temporal evolution of X-ray states for the case of GRO~J1655-40
(1996-1997 outburst) is shown in Fig. 2. The ASM light curve is
displayed in the top panel, while the unabsorbed flux derived from
spectral fits to {\it RXTE} pointed observations are shown in the
bottom panel. Here, the X-ray state is also represented via the choice
of plotting symbol: thermal (red x), hard (blue square), steep power-law
(green triangle), and any intermediate type (yellow circle). In this
particular case, the two intermediate cases exhibit properties that lie
between the thermal and SPL definitions.

The state assignments utilize spectral parameters obtained via the
modeling prescriptions of Sobczak et~al. 1999 \cite{sob99}. There are a few
differences here: all {\it RXTE} observations are considered (adding
programs 10261 and 20187), newer versions of PCA response models are
used, the analyses are restricted to PCU \#2, and a model with broken
power-law (rather than a simple power-law) is used for the beginning
of the outburst, up to and including 1996 June 20. The observations are
grouped into 67 data intervals, combining short observations that
occur on the same day.  However, the last three {\it RXTE} pointings
(beginning 1997 Aug 29) are then 
excluded, since the source flux is below 2 mCrab
and the uncertainties in the spectral parameters is large.
These are very likely to be additional samples of the hard state.
The statistics of the 64 state assignments shown in Fig. 2 are
listed in Table 1.

\begin{figure}
  \includegraphics[height=.75\textheight]{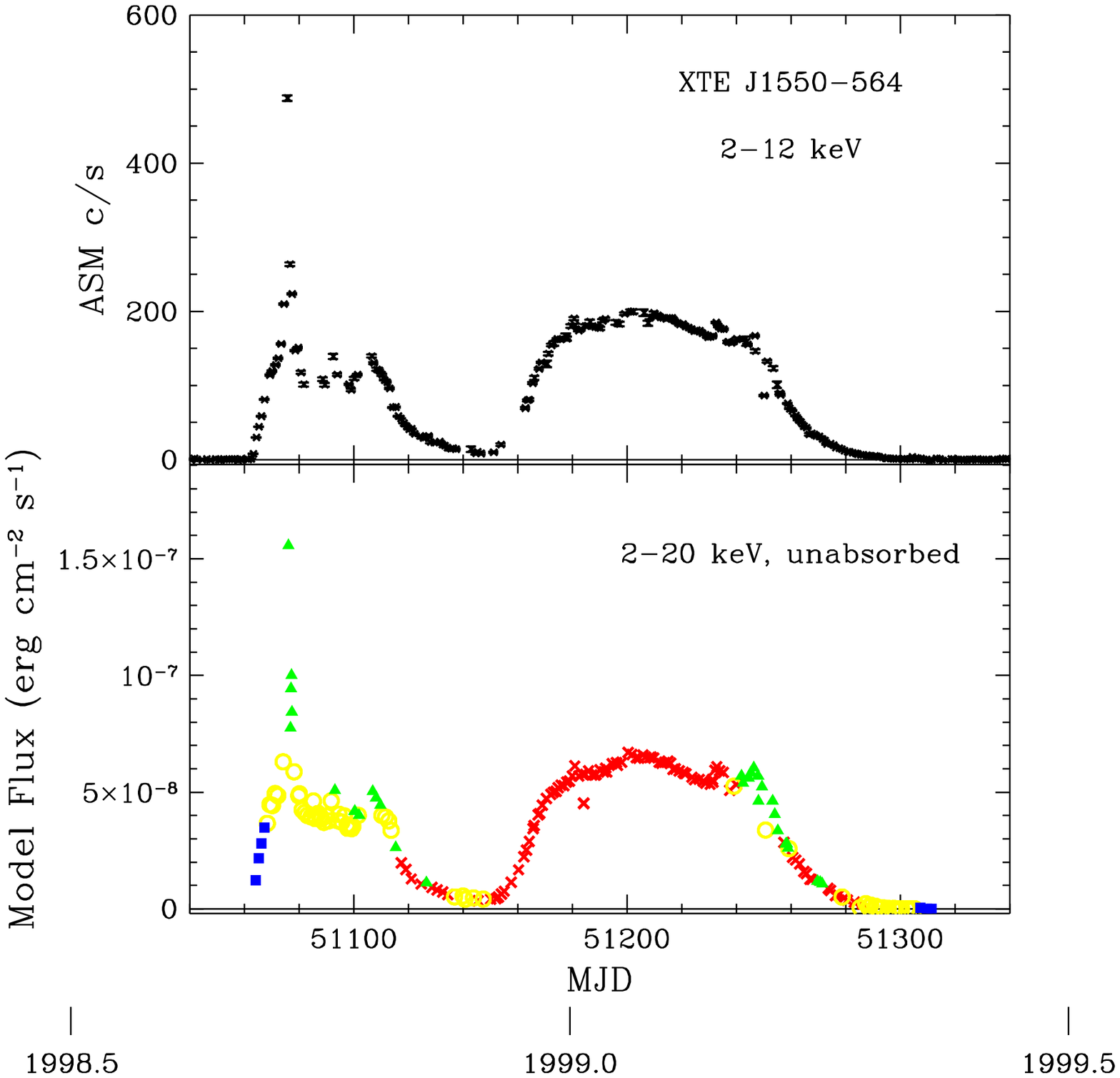}
  \caption{X-ray state evolution during the 1998-1999 outburst of 
  XTE~J1550-564. The top panel shows the ASM light curve.
  The bottom panel shows the model flux from PCA pointed observations,
  again with the X-ray state denoted as: thermal (red x),
  hard (blue square), steep power-law (green triangle), 
  intermediate (yellow circle). }
\end{figure}

The temporal evolution of X-ray states for XTE~J1550-564
(1998-1999 outburst) is shown in Fig. 3, with content and
state representations analogous to Fig. 2.  For this source
spectral modeling efforts follow Sobczak et~al. 2000 \cite{sob00b}.
The re-analysis again targets PCU \#2, a broken power-law model 
is used when it improves the fit significantly,
and this happens frequently for observations before MJD 51140.
Some observations on the same day are grouped together,
with a net of 201 data intervals, and the statistics of state
assignments are included in Table 1. There are a relatively 
large number of intermediate states encountered during this outburst
of XTE~J1550-564, especially in the MJD range 51050--51140.
The latter cases exhibit properties that lie 
between the SPL and hard states (see \cite{mcc05} for more
detailed discussions) and they coincide with the appearance
of ``C'' type QPOs \cite{rem02a}.

The 1998-1999 outburst of XTE~J1550-564 was followed by successively
weaker outbursts in 2000, 2001, 2002, and 2003.  The outburst of 2000
again shows multi-state spectral evolution, but the three weaker
outbursts appear entirely constrained to the hard state (e.g. \cite{bel02}).


\begin{table}
\begin{tabular}{lrrrrr}
\hline
  & \tablehead{1}{r}{b}{Data\\intervals}
  & \tablehead{1}{r}{b}{Thermal\\state}
  & \tablehead{1}{r}{b}{SPL\\state}
  & \tablehead{1}{r}{b}{Hard\\state}
  & \tablehead{1}{r}{b}{Intermediate\\states}   \\
\hline
GRO~J1655-40 (1996-1997)  &  64 &  43 & 16 &  3  &  2 \\
XTE~J1550-564 (1998-1999) & 201 & 102 & 30 &  7  & 62 \\
\hline
\end{tabular}
\caption{Statistics of X-ray State Classifications}
\label{tab:a}
\end{table}


\section{High Frequency QPOs from Black Hole Binaries}
 
High-frequency QPOs (HFQPOs; 40-450 Hz) have been detected thus far in
7 black-hole binaries or candidates.  These are transient subtle
oscillations, with $a \sim 1$\% \cite{mor97,rem99,hom01,str01a,
str01b,rem02,hom03,hom05,rem05}. Frequently, one must select photon
energy bands above 6 keV or above 13 keV in order to gain significant
detection. Furthermore, for statistical reasons most detections
require efforts to group observations with similar spectral and/or
timing characteristics. All of the current detections for black-hole binaries
are all displayed in Fig. 4. HFQPOs above 100 Hz generally occur
during the SPL state \cite{mcc05}.  

\begin{figure}
  \includegraphics[height=.85\textheight]{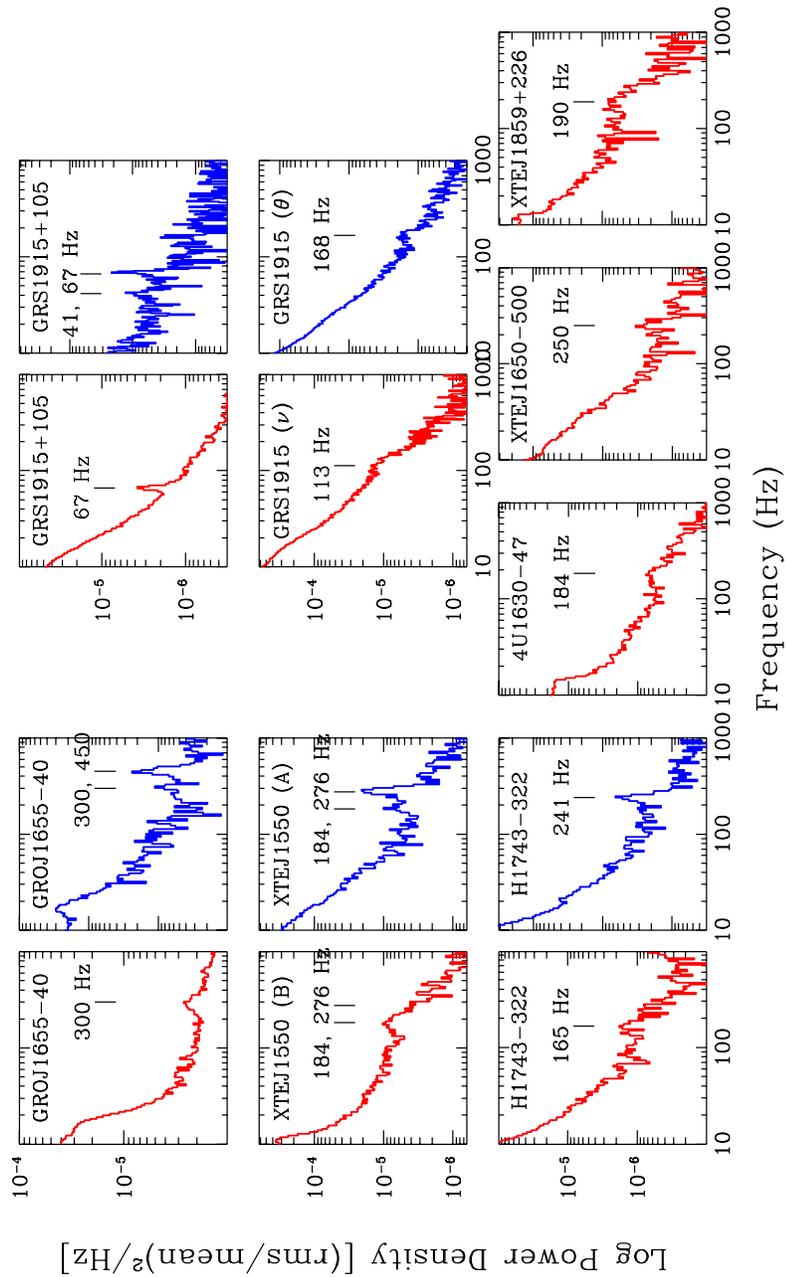}
  \caption{High frequency QPOs (40-450 Hz) seen in 7 black hole binary
  systems.  PDS plotted in red are derived from the energy range 2--30
  keV, while those in blue correspond with 13-30 keV, except for
  H1743-322 (6-30 keV). Four sources show pairs of HFQPOs with
  central frequencies that occur in a 3:2 ratio.  GRS~1915+105 shows
  an additional pair at 41 and 67 Hz.  It should be noted that we
  generally detect one QPO at a time, and QPOs involved in the 3:2
  ratio tend to occur alternately, rather than simultaneously. }
\end{figure}

Four sources (GRO~J1655-40, XTE~J1550-564, GRS~1915+105, and
H1743-322) exhibit pairs of QPOs that have commensurate frequencies in
a 3:2 ratio.  These HFQPOs have frequencies above 100 Hz, and they
generally occur during the SPL state \cite{mcc05}.  GRS~1915+105 
shows an additional, slower QPO pair (41 and 67 Hz) \cite{mor97,str01b} 
that occurs in thermal states with moderate to high luminosity,

In many cases, there is a substantial range in X-ray luminosity within
a set of observations that contribute to a particular HFQPO, or a
particular pair of commensurate HFQPOs in a given source (e.g. \cite{rem02}). 
This supports the conclusion that HFQPO frequency
systems are a stable timing signatures inherent to the accreting black
hole.  This is an important difference from the the kHz QPOs in
neutron-star systems, which exhibit large variations in
frequency.   Moreover, the frequency stability of such a fast
oscillations seems to suggest that HFQPOs may represent an
invaluable means to probe black hole mass and spin via GR theory.

Commensurate HFQPO frequencies can be seen as a signature of an
oscillation driven by some type of resonance condition.  In fact, it
has been proposed by Abramowicz \& Kluzniak \cite{abr01} that QPOs
could represent a resonance in the coordinate frequencies given by
GR for motions around a black hole under strong
gravity (see \cite{mer01}).  Earlier work had used GR coordinate
frequencies and associated beat frequencies to explain QPOs with
variable frequencies in both neutron-star and black-hole systems
\cite{ste99}.

The ``parametric resonance'' concept hypothesizes enhanced emissivity
from accreting matter at a radius where two of the three coordinate
frequencies (i.e. azimuthal, radial, and polar) have commensurate
values that match (either directly or via beat frequencies) the
observed QPOs.  For the cases with known black hole mass, the value of
the dimensionless spin parameter ($a_*$) can be determined
via the application of this resonance model if the correct pair of
coordinate frequencies can be identified.  In fact, reasonable values
($0.25 < a_* < 0.95$) can be derived from the observed HFQPOs for
either 2:1 or 3:1 ratios in either orbital:radial or polar:radial
coordinate frequencies \cite{rem02}.

The driving mechanism that would allow accretion blobs to grow and
survive at the resonance radius has not been specified, and it is
known that there are severe damping forces in the inner accretion disk
\cite{mar98}. On the other hand, ray-tracing calculations under GR 
\cite{sch04} show that the putative blobs could indeed produce 
the HFQPO patterns, and that the choice of $3~\times \nu_0$ versus
$2~\times \nu_0$ for the stronger QPO is governed by the angular width
of the accreting blob.  Clearly, there more work is needed to
investigate this resonance model.

A possible alternative scenario is to extend the models for
``diskoseismic'' oscillations to include non-linear effects that might
drive some type of resonant oscillation.  Diskoseismology treats the
inner disk as a resonance cavity in the Kerr metric
(\cite{kat01,wag99}).  Normal modes have been derived for linear
perturbations, and the extension of this theory would be both very
interesting and difficult. Another alternative is to consider
accretion models that deviate from a thin disk geometry, e.g.
an accretion torus and its oscillation modes under GR \cite{rez03}.

For three systems that show HFQPO pairs with frequencies in 3:2
ratio, we have the fortune of black hole mass estimates.  In most
types of GR oscillations, including the coordinate frequencies
discussed above, the oscillation frequency scales with black-hole mass as
$M^{-1}$, with additional dependence on the dimensionless spin parameter
and the possibly the radius where the oscillations originate. 
Surprisingly, these three cases yield a relationship between
HFQPO frequency and black hole mass that is consistent with a simple
$M^{-1}$ relationship \cite{mcc05}: $\nu_0 = 931 M^{-1}$, where
HFQPOs are seen at frequencies of $2~\times \nu_0$ and $3~\times
\nu_0$. This result suggests that these black holes have
similar values of the dimensionless spin parameter.  Furthermore,
these results offer strong encouragement for efforts
to interpret black-hole HFQPOs via GR theory.

\begin{theacknowledgments}
  This work was supported by the NASA contract to MIT for the ASM and
  EDS instruments on {\it RXTE}. Special thanks are extended to Jeff
  McClintock and Jeroen Homan for many helpful discussions.
\end{theacknowledgments}

\end{document}